\documentclass[runningheads]{llncs}
\usepackage{graphicx}
\usepackage{amsmath}
\begin{document}
\title{On the Parameter Estimation in the Schwartz-Smith's Two-Factor Model}
%
%

\authorrunning{K. Binkowski  et al.}
%
\institute{Macquarie University, North Ryde NSW 2109, Australia}

\author{Karol Binkowski\inst{1}\orcidID{0000-0001-9736-9311} \and
Peilun He\inst{2}\orcidID{0000-0002-2740-3390}  \and \newline
Nino Kordzakhia\inst{3}\orcidID{0000-0002-7853-4550}\and \newline
Pavel Shevchenko\inst{4}\orcidID{0000-0001-8104-8716}
}
\authorrunning{K. Binkowski et al.}
%
\institute{Macquarie University, North Ryde NSW 2109, Australia \\
\email{karol.binkowski@mq.edu.au}\and
\email{peilun.he@students.mq.edu.au}\and
\email{nino.kordzakhia@mq.edu.au}\and
\email{pavel.shevchenko@mq.edu.au}
}

\maketitle              
\begin{abstract}
The two unobservable state variables representing the short and long term factors introduced by Schwartz and Smith in~\cite{SS} for risk-neutral pricing of futures contracts  are  modelled as two correlated Ornstein-Uhlenbeck processes. The Kalman Filter (KF) method has been implemented to estimate the “short” and “long” term factors jointly with unknown model parameters. The parameter identification problem arising within the likelihood function in the KF has been addressed by introducing an additional constraint. The obtained model parameter estimates are the conditional Maximum Likelihood Estimators (MLEs) evaluated within the KF. Consistency of the conditional MLEs is studied. The methodology has been tested on simulated data.  

\end{abstract}

\section{Introduction}
Over more than four decades stochastic processes have been used for modelling of commodity futures prices. In early studies, the commodity prices were modelled using a geometric Brownian motion,~\cite{black}. For pricing of commodity derivatives, the mean-reverting processes  were used for the first time in~\cite{Gibson}; they are also known as Ornstein-Uhlenbeck (O-U) processes, primarily introduced in~\cite{OU}  for modelling of velocity process of Brownian particle under friction.  

This work is based on the paper by Schwartz and Smith in~\cite{SS}, where the O-U two-factor model was used for modelling of short and long equilibrium commodity spot price levels. A commodity spot price $S_t$ is modelled as the sum of two unobservable factors $\chi_t$ and $\xi_t$. Both processes $\chi_t$ and $\xi_t$ are represented  as the mean-reverting processes. In the mean-reverting model, when the commodity price is higher than the equilibrium price level, some new suppliers will enter the market and create downward pressure on the prices. Conversely, when the price is lower than the equilibrium price level, some high-cost suppliers will exit the market and put upward pressure on the prices. In the short term, due to these movements, the price fluctuates temporarily, and it will eventually converge to its equilibrium level over the long term.

Kalman's filtering technique for estimation of the state variables using historical futures prices remains popular for its ability to reproduce realistic commodity futures term structure, see e.g.~\cite{ames2016risk}. In~\cite{CortazarGonzalo2016CPFF}, the authors improved the performance of the Kalman Filter by deriving the commodity spot prices from futures prices which have had incorporated an analyst's forecasts of spot prices. In~\cite{ChengBenjamin2018Polc}, the Kalman Filter is used to study the effect of stochastic volatility and interest rates on the commodity spot prices using the market prices of long-dated futures and options. The Kalman technique was used in~\cite{pavel} for calibration and filtering of partially observable processes using particle Markov chain Monte Carlo approach. The authors of~\cite{ewald} developed the extended Kalman Filter for estimation of the state variables in the two-factor model for the commodity spot price and its yield developed in~\cite{schwartz1997stochastic}.

Our motivation is driven by the fact that the parameter estimation problem in the linear system using the Kalman Filter cannot be overlooked whilst the estimation of the state variables remains the priority.   In the different setup the parameter estimation problem for bivariate O-U process using Kalman Filter has been studied in~\cite{favetto2010parameter,KutoyantsYuryA.2019Opeo}. 

In this paper, we conduct the simulation study where the Kalman Filter is deployed for estimation of the model parameters jointly with the components of the logarithm of spot price processs, which is used in the pricing formula for the futures contracts.  
In Section 2, we provide the notation and analytical formulae used for the formulation of linear partially observable system specific for commodity futures pricing given in~\cite{SS}. In Section 3, we present the details of our implementation of KF algorithm designed for estimation of the parameters of the multi-dimensional partially observable linear system jointly with the estimation of unobserved state variables $\chi_t$ and $\xi_t$. The results of the simulation study are presented in Section 4.

\section{Two-Factor Model} 
In this section, we provide a brief description of Schwartz-Smith's model from~\cite{SS}. We discuss the risk-neutral
setup used for futures pricing and present
the formulae for futures prices, incorporating the short and long term dynamic factors.

\subsection{A Commodity Spot Price Modelling}
We model the logarithm of the spot price $S_t$ using the additive
model, 
\[\log(S_t) = \chi_t + \xi_t,\] where \(\chi_t\)  and \(\xi_t\) are the short and long dynamic factors, respectively. We assume that changes in \(\chi_{t}\) are temporary and it
converges to 0 in a long term, following an O-U process
\[d\chi_{t} = -\kappa\chi_{t}dt + \sigma_{\chi}dZ_{t}^{\chi},\] 
with the mean 0. The
changes in the equilibrium level of \(\xi_t\) are expected to persist and
the process itself is assumed to be mean-reverting
\[d\xi_{t} = (\mu_{\xi} - \gamma\xi_{t})dt + \sigma_{\xi}dZ_{t}^{\xi}.\]
The processes \(Z_{t}^{\chi}\) and \(Z_{t}^{\xi}\) are correlated standard Brownian motions with \\
\(dZ_{t}^{\chi}dZ_{t}^{\xi} = \rho_{\chi\xi}dt\), and $ \rho =\rho_{\chi\xi}$. Given the initial values \(\chi_0\) and \(\xi_0\),
\(\chi_t\) and \(\xi_t\) are jointly normally distributed with expected value
\begin{equation}
E[(\chi_t, \xi_t)] = (e^{-\kappa t}\chi_0, \frac{\mu_\xi}{\gamma}(1 - e^{-\gamma t}) + e^{-\gamma t}\xi_0), t \ge 0
\label{eq:meanX}
\end{equation}
and covariance matrix
\begin{equation}
Cov[(\chi_t, \xi_t)] = \left(\begin{array}{cc}
\frac{1 - e^{-2\kappa t}}{2\kappa}\sigma_{\chi}^2 & \frac{1 - e^{-(\kappa + \gamma) t}}{\kappa + \gamma}\sigma_{\chi}\sigma_{\xi}\rho_{\chi\xi} \\
\frac{1 - e^{-(\kappa + \gamma) t}}{\kappa + \gamma}\sigma_{\chi}\sigma_{\xi}\rho_{\chi\xi} & \frac{1 - e^{-2\gamma t}}{2\gamma}\sigma_{\xi}^2
\end{array}\right).
\label{eq:covX}
\end{equation}
Derivations of \eqref{eq:meanX} and \eqref{eq:covX} are given in Appendix
\ref{rmd-derivation}. Therefore, the logarithm of the spot price is normally
distributed with mean
\[E[\log(S_{t})] = e^{-\kappa t}\chi_0 + \frac{\mu_\xi}{\gamma}(1 - e^{-\gamma t}) + e^{-\gamma t}\xi_0\]
and variance
\[Var[\log(S_{t})] = \frac{1 - e^{-2\kappa t}}{2\kappa}\sigma_{\chi}^2 + \frac{1 - e^{-2\gamma t}}{2\gamma}\sigma_{\xi}^2 + 2\frac{1 - e^{-(\kappa + \gamma) t}}{\kappa + \gamma}\sigma_{\chi}\sigma_{\xi}\rho_{\chi\xi}.\]
Hence \(S_t\), the commodity spot price, is log-normally distributed and
\[E(S_{t}) = \exp(E[\log(S_{t})] + \frac{1}{2}Var[\log(S_{t})]), \] or
\[\log[E(S_{t})] = e^{-\kappa t}\chi_0 + \frac{\mu_\xi}{\gamma}(1 - e^{-\gamma t}) + e^{-\gamma t}\xi_0 + \]
\begin{equation}
\frac{1}{2}\left(\frac{1 - e^{-2\kappa t}}{2\kappa}\sigma_{\chi}^2 + \frac{1 - e^{-2\gamma t}}{2\gamma}\sigma_{\xi}^2 + 2\frac{1 - e^{-(\kappa + \gamma) t}}{\kappa + \gamma}\sigma_{\chi}\sigma_{\xi}\rho_{\chi\xi}\right),
\label{eq:logES}
\end{equation}
where \(\frac{\mu_{\xi}}{\gamma}\) is the mean parameter, \(\sigma_{\chi}\) and \(\sigma_{\xi}\)
are the volatilities, \(\gamma\) and \(\kappa\) are so-called the speed of
mean-reversion parameters of \(\chi\) and \(\xi\) processes,
respectively.

\subsection{Risk-Neutral Approach to Spot Price Modelling}

In this section we introduce two additional parameters which can be interpreted as the market price of commodity spot price risk. The approach stems from the risk-neutral pricing theory for futures, developed in~\cite{black}. Hence, the Schwartz-Smith's model with additional parameters can be rewritten as follows
\[d\chi_{t} = (-\kappa\chi_{t} - \lambda_{\chi})dt + \sigma_{\chi}dZ_{t}^{\chi^*}, \]
\[d\xi_{t} = (\mu_{\xi} - \gamma\xi_{t}-\lambda_{\xi} )dt + \sigma_{\xi}dZ_{t}^{\xi^*}, \]
where the parameters \(\lambda_{\chi}\) and \(\lambda_{\xi}\) appear as the risk-neutral mean correction terms. Under the risk-neutral measure, \(\chi_t\) and \(\xi_t\)
are also jointly normally distributed with mean
\[E^*[(\chi_t, \xi_t)] = (e^{-\kappa t}\chi_0 - \frac{\lambda_{\chi}}{\kappa}(1 - e^{-\kappa t}), \frac{\mu_{\xi} - \lambda_{\xi}}{\gamma}(1 - e^{-\gamma t}) + e^{-\gamma t}\xi_0)\]
and covariance matrix \[Cov[(\chi_t, \xi_t)]^* = Cov[(\chi_t, \xi_t)].\]
The logarithm of commodity spot price is normally distributed with mean
\[E^*[\log(S_t)] = e^{-\kappa t}\chi_0 - \frac{\lambda_{\chi}}{\kappa}(1 - e^{-\kappa t}) + \frac{\mu_{\xi} - \lambda_{\xi}}{\gamma}(1 - e^{-\gamma t}) + e^{-\gamma t}\xi_0\]
and variance \[Var^*[\log(S_t)] = Var[\log(S_t)].\] The spot price is
log-normally distributed with
\begin{equation}
\log[E^*(S_t)] = E^*[\log(S_t)] + \frac{1}{2}Var^*[\log(S_t)] = e^{-\kappa t}\chi_0 + e^{-\gamma t}\xi_0 + A(t), 
\label{eq:logESRN}
\end{equation}
where
\[A(t) = -\frac{\lambda_{\chi}}{\kappa}(1 - e^{-\kappa t}) + \frac{\mu_{\xi} - \lambda_{\xi}}{\gamma}(1 - e^{-\gamma t}) + \]
\[\frac{1}{2}\left(\frac{1 - e^{-2\kappa t}}{2\kappa}\sigma_{\chi}^2 + \frac{1 - e^{-2\gamma t}}{2\gamma}\sigma_{\xi}^2 + 2\frac{1 - e^{-(\kappa + \gamma)t}}{\kappa + \gamma}\sigma_{\chi}\sigma_{\xi}\rho_{\chi\xi}\right).\]
In \eqref{eq:logESRN} the parameters \(\lambda_{\chi}\) and
\(\lambda_{\xi}\) appear due to the adjustment made in
\eqref{eq:logES}.

\subsection{Risk-Neutral Approach to Pricing of Futures}\label{pricing-of-futures-contracts}

A futures contract is defined as an agreement to
trade or own an asset in the future,~\cite{hull}. We are interested to
know what is the price of such contract at present. Let \(F_{0, T}\) be
the current market price of the futures contract with maturity \(T\).
For elimination of arbitrage, colloquially known as a ``free-lunch", the futures price must be equal to the expected commodity spot price
at its delivery time \(T\). Hence, under the risk-neutral measure
from Section 2.2, assuming  zero interest rate, we obtain 
\[\log(F_{0, T}) = \log[E^*(S_T)] = e^{-\kappa T}\chi_0 + e^{-\gamma T}\xi_0 + A(T).\]
We denote 
\[x_t = \left(\begin{matrix} \chi_t \\ \xi_t \end{matrix} \right),\;\; c = \left(\begin{array}{c} 0 \\ \frac{\mu_{\xi}}{\gamma}(1 - e^{-\gamma \Delta t}) \end{array}\right),\;\; G = \left(\begin{array}{cc}e^{-\kappa \Delta t} & 0 \\ 0 & e^{-\gamma \Delta t} \end{array}\right), \]
$\Delta t$ is the discretization width. \\
Let \(w_t\) be a column-vector of normally distributed random variables, 
\[
E(w_t) = 0
\] 
and  
\[
W = Cov(w_t) = Cov[(\chi_{\Delta t}, \xi_{\Delta t})].
\] 
In discrete time, we will obtain the following AR(1) dynamics for the bivariate state
variable \(x_t\)
\begin{equation}
x_t = c + Gx_{t - 1} + w_t.
\label{eq:xt}
\end{equation}
The
relationship between \(x_t\) and the observed futures prices
is given by
\begin{equation}
y_t = d_t + F_t'x_t + v_t,
\label{eq:yt}
\end{equation}
where \[y_t' = (\log(F_{T_1}), \log(F_{T_2}), \dots, \log(F_{T_n})),\]
\[d_t' = (A(T_1), A(T_2), \dots), A(T_n),\]
\[F_t = \begin{pmatrix} e^{-\kappa (T_1-t)},  e^{-\kappa (T_2-t)}, \dots, e^{-\kappa (T_n-t)} \\ e^{-\gamma (T_1-t)}, e^{-\gamma (T_2-t)}, \dots, e^{-\gamma (T_n-t)} \end{pmatrix},\]
and \(v_t\) is a \(n \times 1\) vector of independent,  normally
distributed random variables \(E(v_t) = 0\) and \(Cov(v_t) = V\) and $T_1,T_2,...,T_n$ are the futures maturity times. 
We assume that \(V\) is a diagonal matrix with the vector \(v=(s_1^2, s_2^2, \dots, s_n^2)\) of non-zero elements
 on the diagonal. The number of the
futures contracts is \(n\). Let \(\mathcal{F}_t\) be a $\sigma$-algebra
generated by the futures contracts up to time \(t\) and $\theta = (\kappa, \gamma, \mu_{\xi}, \sigma_{\chi}, \sigma_{\xi}, \rho, \lambda_{\chi}, \lambda_{\xi}, v)$ be a vector of unknown parameters. The model \eqref{eq:xt} - \eqref{eq:yt} is similar to the model discussed in Chapter 3,~\cite{harvey}.

The conditional log-likelihood
function of $y=(y_1, y_2, \dots, y_{n_T})$ is \[l(\theta; y) = \sum_{t = 1}^{n_T}p(y_t|\mathcal{F}_{t - 1}), \] where \(p(y_t|\mathcal{F}_{t - 1})\) is the probability density of
\(y_t\) given the information available until $t-1=t-\Delta$ and $n_T$ is the number of time-points. We
assume that the prediction errors
\(e_t = y_t - E(y_t|\mathcal{F}_{t - 1})\) have a multivariate normal distribution, then the log-likelihood function is
\begin{equation}
l(\theta; y) = -\frac{nn_T\log(2\pi)}{2} - \frac{1}{2}\sum_{t = 1}^{n_T}[\log(\det(L_{t | t - 1})) + e_t'L_{t | t - 1}^{-1}e_t]
\label{eq:loglikelihood}
\end{equation}
where \(L_{t | t - 1} = Cov(e_t|\mathcal{F}_{t - 1})\). The vector of unknown
parameters $\theta$ will be estimated by maximising the log-likelihood function from \eqref{eq:loglikelihood}. However, the maximisation of \(l(\theta; y) \) is inhibited by the parameter identification problem. This fact can be proved analyticaly by mathematical induction since the prediction error $e_t$  
and covariance matrix $L_{t|t-1}$ are invariant to label switching of coordinates of $x_t$.

\section{Kalman Filter}
In this section, we are using Kalman Filter to estimate the unobservable
vector of state variables \(x_t = (\chi_t, \xi_t)'\) using simulated
\(y_t\). We recall the equations \eqref{eq:xt} and \eqref{eq:yt} for \(x_t\)
and \(y_t\), respectively

\[x_t = c + Gx_{t - 1} + w_t,\] \[y_t = d_t + F_t'x_t + v_t.\]

For initialisation of the Kalman Filter we use the expectation and covariance matrix, suggested in~\cite{karol}
\[a_0 = E(x_0) = \left(0, \frac{\mu_{\xi}}{\gamma}\right)'\]
and
\[P_0 =Cov(x_0) =  \left(\begin{array}{cc} \frac{\sigma_{\chi}^2}{2\kappa} & \frac{\sigma_{\chi}\sigma_{\xi}\rho_{\chi \xi}}{\kappa + \gamma} \\ \frac{\sigma_{\chi}\sigma_{\xi}\rho_{\chi \xi}}{\kappa + \gamma} & \frac{\sigma_{\xi}^2}{2\gamma} \end{array}\right).\]
The flowchart for Kalman Filter is given in Figure 1. The recursive process is constructed by starting  at  \(x_0 \sim N(a_0, P_0)\).

\begin{figure}
	\centering 
	\includegraphics[scale=0.6]{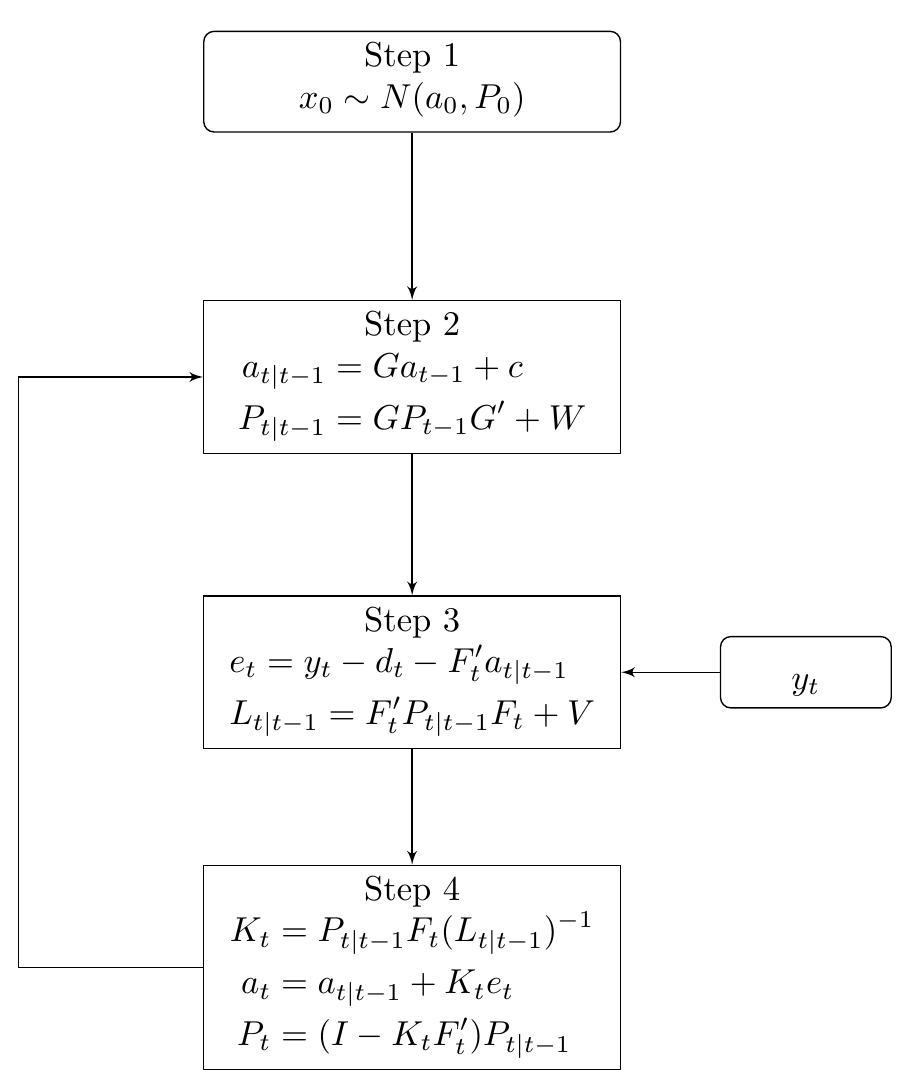} \caption{Flowchart for Kalman Filter.}\label{fig1}
\end{figure}
\noindent Next, we will evaluate the conditional expected value  $a_{t|t-1}$ and the conditional covariance matrix $P_{t|t-1}$ of the state vector $x_t$. At Step 3 the new observation $y_t$ is entered and we calculate the prediction error $e_t$ and covariance matrix $L_{t|t-1}$. Next, we update $a_t$ and $P_t$ through the Kalman's gain matrix $K$. Finally, we calculate the log-likelihood function at time \(t\)
\[l_t = -\log(\det(L_{t|t - 1})) - e_t'L_{t|t-1}^{-1}e_t.\]
We complete the recursive process from \(t = 1\) to \(n_T\) by
summing  up all \(l_t\)'s  to obtain  the log-likelihood function
$$l(\theta;y) = \sum_{t=1}^{n_T} l_t. $$
Then we maximise \(l(\theta;y)\) for obtaining the conditional maximum likelihood estimate (MLE) of \(\theta\).

\noindent The overview of R packages for Kalman Filter is given in~\cite{packager}. The R packages, which were adapted to this study DSE~\cite{gilbert2005brief}, KFAS~\cite{helske2016kfas} and ASTSA~\cite{shumway2017time}, were mostly sensitive to the choice of the initial values of the model parameters.

\section{Simulation Study}
In this section, we present the results of the simulation study conducted for purposes of validating the use of Kalman Filter for estimation of the state vector  $x_t$ jointly with the model parameters $\theta$. The simulation study has been programmed as follows.
\begin{itemize}
\item [1.] Set  $\theta = (\kappa, \gamma, \mu, \sigma_{\chi}, \sigma_{\xi}, \rho, v)$ as the  vector of  true values. 

\item [2.]  Simulate $x_t$ and $y_t$ using the true values of parameters set in $\theta$. 

\item [3.] Set the intervals for searching of unknown parameters. Locate the grid over the Cartesian product of these intervals. 

\item [4.]  Do grid-search for finding the ``best" initial vector $\theta_0.$ 

\item [5.]  Maximise the log-likelihood function \(l(\theta;y)\) using the ``best" initial vector  $\theta_0$. For circumventing the parameter identification problem in $x$ vector, we added the constraint $\kappa \ge \gamma$. In Schwartz-Smith's model, the speed of mean-reversion parameter $\gamma$ of the long term factor $\xi_t$ is natrually dominated by $\kappa$, the speed of mean-reversion of the short term factor $\chi_t$. 

Obtain  $\hat{\theta}$, the MLE of $\theta$. 

\item [6.]  At $\hat{\theta}$ obtain the estimates of the state variables $\hat{\chi}$ and $\hat{\xi}$. 




\end{itemize}
The grid search for the ``best" initial set of the parameters' values allowed to overcome the problem of sensitivity to the initial values. 

\noindent Further, for simplicity we assume  $\lambda_{\chi}= \lambda_{\xi}=0 $ and $s_1^2=s_2^2=...=s_n^2=s^2$. 

\noindent The model parameter estimates obtained by using the above procedure (1-6) are presented in Table \ref{tab1}.


\begin{table}
	\caption{$\hat{\theta}$ for \(n \in [500, 8000]\); NLL stands for \(-l(\theta;y).\)    }\label{tab1}
	\centering
	\begin{tabular}{|l|l|l|l|l|l|l|l|l|}
		\hline
		$n$ &  $\kappa$ & $\gamma$ & $\mu$ & $\sigma_{\chi}$ & $\sigma_{\xi}$ & $\rho$ & $s$ & NLL \\
		\hline
		\hline
		500  &  1.2775 & 0.0350 & -0.1037 & 1.3910 & 0.1811 & -0.9517 & 0.0301 & -12805  \\
		1000 &  1.4973 & 0.9896 & -2.0078 & 1.1990 & 0.5409 & -0.3623 & 0.0299 & -25612 \\
		2000 &  1.5327 & 0.9834 & -1.9932 & 1.2579 & 0.4351 & -0.3604 & 0.0297 & -51376 \\
		4000 &  1.4895 & 1.0139 & -2.0361 & 1.3376 & 0.4736 & -0.5880 & 0.0300 & -102549 \\
		6000 & 1.4711 & 0.9913 & -1.9931 & 1.3526 & 0.4261 & -0.6439 & 0.0300 & -153904 \\
		8000 & 1.4938 & 0.9960 & -2.0008 & 1.3198 & 0.3936 & -0.6078 & 0.0300 & -205120 \\
		\hline
		True ($\theta_0$) &  1.50 & 1.00 & -2.00 & 1.30 & 0.30 & -0.70 & 0.03 &  \\
		\hline
	\end{tabular}
\end{table}

\noindent For some sample sizes $n$ from the range $(500, 8000)$, the ``best" initial values are given in Table \ref{tab2}.  These initial values were used for obtaining the corresponding optimal model parameter estimates in Table \ref{tab1}. 

\newpage

\begin{table}[ht]
	\caption{``Best" initial values for $\hat{\theta}$. }\label{tab2}
	\centering
		\begin{tabular}{|l|l|l|l|l|l|l|l|}
			\hline
			$n$ &  $\kappa$ & $\gamma$ & $\mu$ & $\sigma_{\chi}$ & $\sigma_{\xi}$ & $\rho$ & $s$ \\
			\hline
			\hline
			500  &  2.2525 & 0.7575 & 1.7500 & 1.5025 & 1.0050 & -0.5000 & 0.5000 \\
			1000 &  1.5050 & 0.7575 & -0.5000 & 1.0050 & 1.5025 & -0.5000 & 0.7500 \\
			2000 & 0.7575 & 0.7575 & -2.7500 & 1.5025 & 1.5025 & 0.5000 & 0.5000 \\
			4000 & 2.2525 & 0.7575 & 1.7500 & 1.0050 & 1.5025 & 0.5000 & 0.2500 \\
			6000 & 1.5050 & 1.5050 & -0.5000 & 0.5075 & 1.5025 & 0.5000 & 0.2500 \\
			8000 & 2.2525 & 0.7575 & -0.5000 & 0.5075 & 1.0050 & -0.5000 & 0.2500 \\
			\hline
	\end{tabular}
\end{table}

\begin{figure}
	\includegraphics[width=\textwidth]{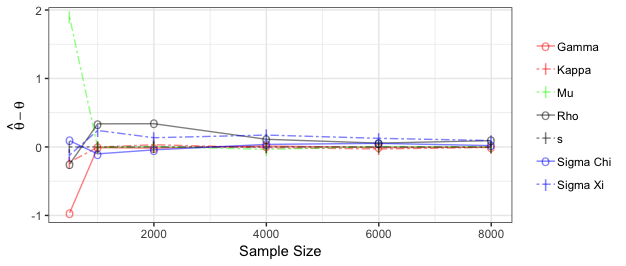}
	\caption{ Componentwise estimation error plots for  $\theta $ versus  $n$.} 
	\label{fig2}
\end{figure}

\noindent The convergence of the parameter estimates can be seen in Figure \ref{fig2}, where the estimation errors $\hat{\theta}_i - \theta_{i}, i=1,2,...,7$ are plotted versus the sample size $n$, $\theta$ is the vector of true parameter values. 

\begin{figure}
	\centering
	\includegraphics[width=\textwidth]{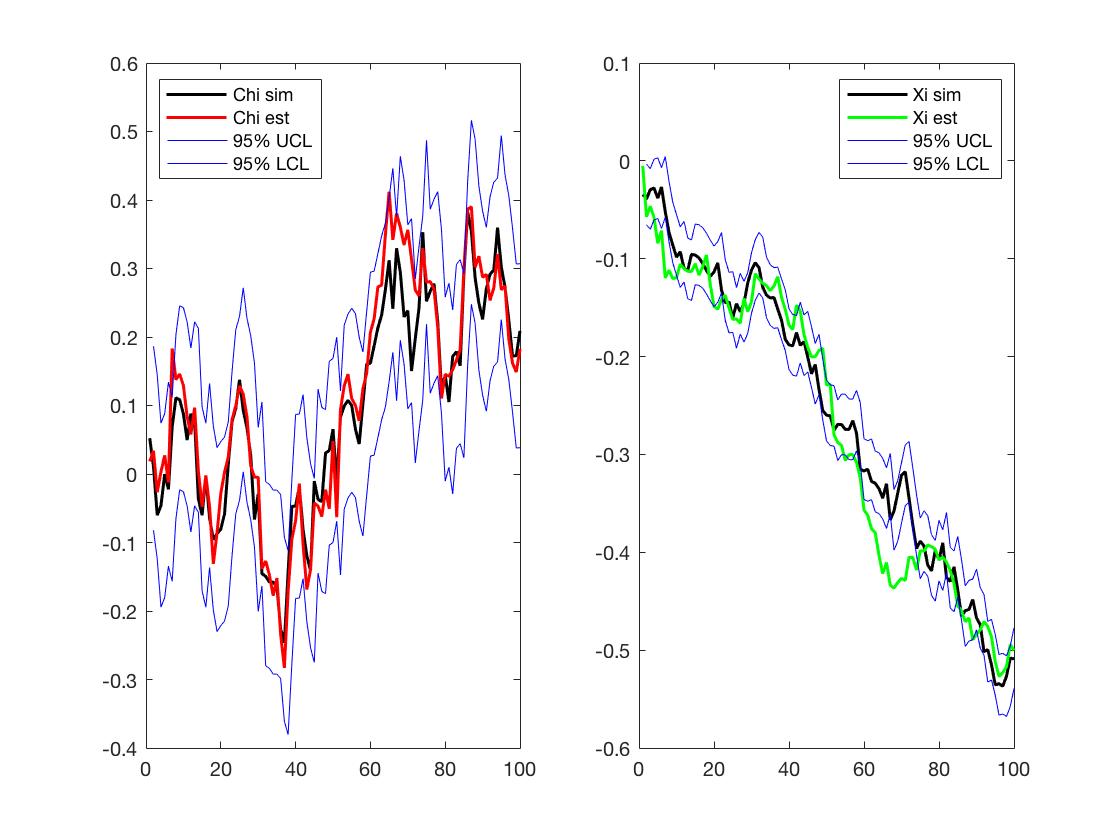}
	\caption{Estimated $(n=8000), {\hat\chi}_t$, ${\hat\xi}_t$ and simulated $\chi_t$, $\xi_t$ paths  with their 95\% CI.} 
	\label{fig3}
\end{figure}

\noindent The paths of the estimated state variables ${\hat\chi}$ and ${\hat\xi}$ were obtained through Kalman Filter along with the simulated trajectories of $\chi$ and $\xi$ and their 95\%-Confidence Intervals (CIs) based on the true  values of the model parameters are presented in Figure \ref{fig3}. 


\newpage

\begin{figure}[ht]
	\centering
	\includegraphics[width=\textwidth]{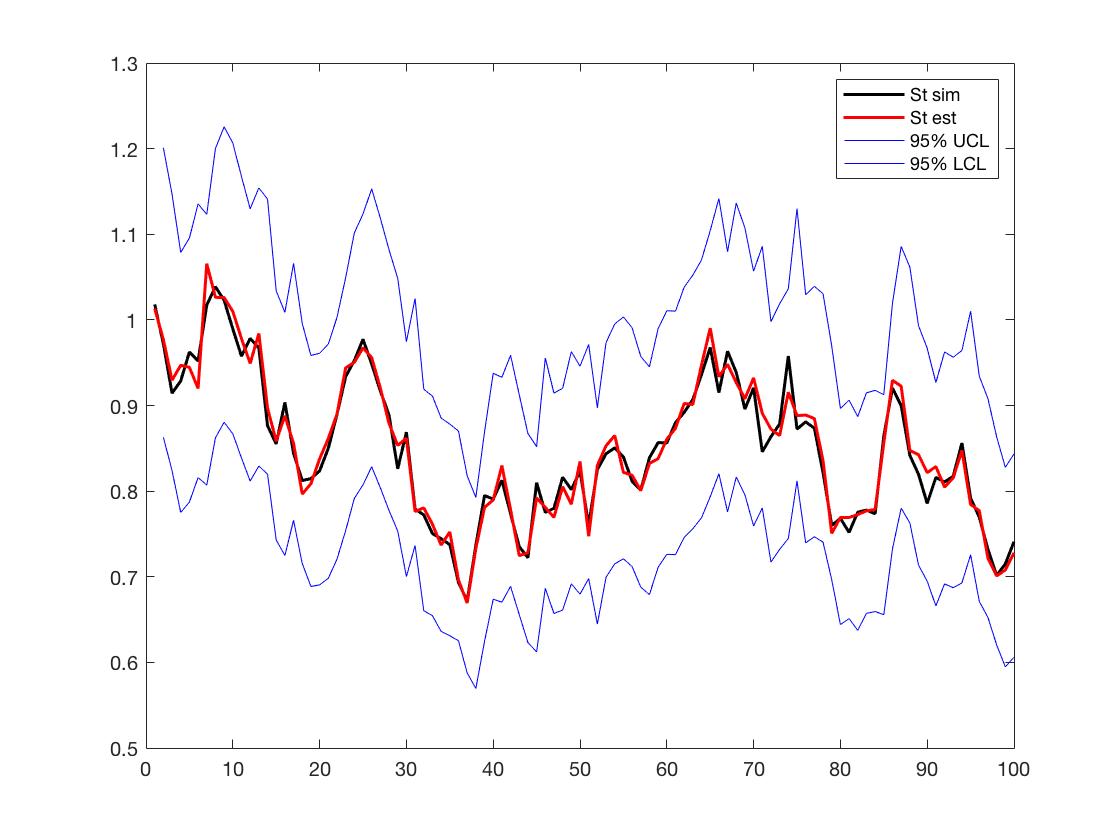}
	\caption{Estimated $\hat S_t$ and simulated $S_t$ with its 95\% CI.} 
	\label{fig4}
\end{figure}

The plots of the paths of the estimated  $\hat S_t = \exp(\hat\chi_t+\hat\xi_t)$ and the spot prices $S_t = \exp(\chi_t+\xi_t)$ computed using the simulated paths of \({\chi} \) and \({\xi}\)       along with 95\% CI based on the true model parameter values $\theta$ are presented in Figure \ref{fig4}.

In the AWS Australian Sydney computing center, for computations in Matlab we used c5.18xlarge instances (CPU 36) taking 10 hours for \(n=8000\).  

\newpage

\section{Conclusions}
In this paper, the parameter estimation problem has been studied in the linear partially observable system, which is specific for commodity futures prices developed in  the Schwartz-Smith's two-factor model in the risk-neutral setting. 
In the simulation study, the Kalman Filter algorithm has been implemented and tested for estimation of the parameters of the bivariate O-U process $x$ jointly with the estimation of its unobserved components $\chi$ and $\xi$. In this study, we suggested the remedy for rectifying the parameter identification problem arising within MLE procedure in the Kalman Filter. The simulation study illustrated the robustness of the grid-search and consistency of the estimates of the model parameters and state vector $x$.

\newpage

\appendix

\section{Derivations of \eqref{eq:meanX} and
	\eqref{eq:covX}}\label{rmd-derivation}

In Section 2.1, we define an Ornstein-Uhlenbeck process as
\begin{equation}
d\chi_{t} = -\kappa\chi_{t}dt + \sigma_{\chi}dZ_{t}^{\chi}
\label{eq:aone}
\end{equation}
and a mean reverting process
\begin{equation}
d\xi_{t} = (\mu_{\xi} - \gamma\xi_{t})dt + \sigma_{\xi}dZ_{t}^{\xi},
\label{eq:atwo}
\end{equation}
where $dZ_t^{\chi}, dZ_t^{\xi} \sim N(0, \sqrt{\Delta t})$ are correlated standard Brownian motions. Here we provide details to get equation \eqref{eq:meanX} amd \eqref{eq:covX}
from \eqref{eq:aone} and \eqref{eq:atwo}. Firstly, from \eqref{eq:aone},

\[\Delta \chi_t = -\kappa \chi_t \Delta t + \sigma_{\chi} \sqrt{\Delta t} \epsilon_{\chi}.\]
Therefore,
\begin{equation}
\chi_{t+1} = (1 - \kappa \Delta t) \chi_t + \sigma_{\chi} \sqrt{\Delta t} \epsilon_{\chi}.
\label{eq:athree}
\end{equation}
Similarly, from \eqref{eq:atwo}, we get
\begin{equation} 
\xi_{t+1} = (1 - \gamma \Delta t) \xi_t + \mu_{\xi}\Delta t + \sigma_{\xi} \sqrt{\Delta t} \epsilon_{\xi}, 
\label{eq:afour}
\end{equation}
where \(\epsilon_{\chi}, \epsilon_{\xi} \sim N(0,1)\). Let
\(Corr(\epsilon_{\chi}, \epsilon_{\xi}) = \rho\) and $w = \left( \begin{matrix} \sigma_{\chi} \sqrt{\Delta t} \epsilon_{\chi} \\ \sigma_{\xi} \sqrt{\Delta t} \epsilon_{\xi}\end{matrix} \right)$, then
\begin{equation}
W=Var(w) = \left(\begin{matrix} \sigma_{\chi}^2 \Delta t & \rho \sigma_{\chi} \sigma_{\xi} \Delta t \\ \rho \sigma_{\chi} \sigma_{\xi} \Delta t & \sigma_{\xi}^2 \Delta t \end{matrix} \right).
\label{eq:a12}
\end{equation}
Let \(X_t = \left(\begin{matrix} \chi_t \\ \xi_t \end{matrix} \right)\),
\(c = \left(\begin{matrix} 0 \\ \mu_{\xi}\Delta t \end{matrix} \right)\) and 
\(G = \left(\begin{matrix} 1 - \kappa \Delta t & 0 \\ 0 & 1 - \gamma \Delta t \end{matrix} \right)\). 
Then from \eqref{eq:athree} and \eqref{eq:afour} we get
\[X_{t+1} = c + GX_t + w_{t+1}.\] 
Let \(\phi = 1 - \kappa \Delta t\),
\(\psi = 1 - \gamma \Delta t\). Then
$$E(X_t) = \left(\begin{matrix} (1-\kappa \Delta t) \chi_{t-1} \\ (1 - \gamma \Delta t) \xi_{t-1} + \mu_{\xi} \Delta t \end{matrix} \right)  $$
$$ = \left( \begin{matrix} (1-\kappa \Delta t)^n \chi_0 \\ (1 - \gamma \Delta t)^n \xi_0 + (1-\gamma \Delta t)^{n-1} \mu_{\xi} \Delta t + ... + (1-\gamma \Delta t)^0 \mu_\xi \Delta t \end{matrix} \right)$$

$$ = \left(\begin{matrix} \phi^n \chi_0 \\ \psi^n \xi_0 + \mu_{\xi} \Delta t \frac{1-(1-\gamma \Delta t)^n}{\gamma \Delta t} \end{matrix} \right)$$ 

\begin{equation} 
= \left( \begin{matrix} \phi^n \chi_0 \\ \psi^n \xi_0 + \frac{\mu_{\xi}}{\gamma}(1-\psi^n) \end{matrix} \right), 
\label{eq:afive}
\end{equation}

and 

$$Var(X_t) = G\cdot Var(X_{t-1}) \cdot G' + W = G^{n}Var(X_0)(G')^{n} + G^{n-1}W(G')^{n-1} + ... + G^0W(G')^0.$$
If we assume $Var(X_0) = 0$, we can get 
$$Var(X_t) = G^{n-1}W(G')^{n-1} + ... + G^0W(G')^0$$
$$= \left(\begin{matrix} \sigma_{\chi}^2 \Delta t \sum_{i=0}^{n-1}\phi^{2i} & \rho \sigma_{\chi} \sigma_{\xi} \Delta t \sum_{i=0}^{n-1}(\phi\psi)^{i} \\ \rho \sigma_{\chi} \sigma_{\xi} \Delta t \sum_{i=0}^{n-1}(\phi\psi)^{i} & \sigma_{\xi}^2 \Delta t \sum_{i=0}^{n-1}\psi^{2i} \end{matrix} \right) $$
\begin{equation}
=\left(\begin{matrix} \sigma_{\chi}^2 \Delta t \frac{1-\phi^{2n}}{1-\phi^2} & \rho \sigma_{\chi} \sigma_{\xi} \Delta t \frac{1-(\phi \psi)^n}{1-\phi \psi} \\ \rho \sigma_{\chi} \sigma_{\xi} \Delta t \frac{1-(\phi \psi)^n}{1-\phi \psi} & \sigma_{\xi}^2 \Delta t \frac{1-\psi^{2n}}{1-\psi^2} \end{matrix} \right). 
\label{eq:asix}
\end{equation}

When \(n \to \infty\), \(\Delta t = t / n \to 0\),
\(\Delta t^2 = 0\), then
$$\phi^n = (1-\frac{\kappa t}{n})^n \to e^{-\kappa t}, $$
$$\psi^n = (1-\frac{\gamma t}{n})^n \to e^{-\gamma t}, $$
$$(\phi\psi) ^ n = (1-(\kappa + \gamma) t / n)^n \to e^{-(\kappa + \gamma) t}, $$
$$ 1-\phi^2 = 2\kappa \Delta t, 1-\psi^2 = 2\gamma \Delta t, 1 - \phi \psi = (\kappa + \gamma)\Delta t. $$
From equation \eqref{eq:afive} and \eqref{eq:asix}, we have 
\[E(X_t) = \left(\begin{matrix} e^{-\kappa t} \chi_0 \\ e^{-\gamma t} \xi_0 +  \frac{\mu_{\xi}}{\gamma}(1 - e^{-\gamma t}) \end{matrix} \right)\]
and 
\[Var(X_t) = \left(\begin{matrix} \frac{\sigma_{\chi}^2}{2\kappa}(1-e^{-2\kappa t}) & \frac{\rho \sigma_{\chi} \sigma_{\xi}}{\kappa + \gamma} (1-e^{-(\kappa + \gamma) t}) \\ \frac{\rho \sigma_{\chi} \sigma_{\xi}}{\kappa + \gamma} (1-e^{-(\kappa + \gamma) t}) & \frac{\sigma_{\xi}^2}{2\gamma}(1-e^{-2\gamma t}) \end{matrix} \right), \]
in the linearised form $Var(X_{{\Delta} t})\approx W$ from \eqref{eq:a12}.

%
%
%
\nocite{*}
\bibliographystyle{splncs04}
\bibliography{Reference}

\end{document}